# Fluorescence-Enhanced Mid-Infrared Photothermal Microscopy

Yi Zhang[#][1], Haonan Zong[#][2], Cheng Zong[2], Yuying Tan[3], Meng Zhang[2], Yuewei Zhan[3], Ji-Xin Cheng[*][1][2][3][4]

[1]Department of Physics, Boston University, Boston, MA 02215, USA
[2]Department of Electrical and Computer Engineering, Boston University, Boston, MA 02215, USA
[3]Department of Biomedical Engineering, Boston University, Boston, MA 02215, USA
[4]Photonics Center, Boston University, Boston, MA 02215, USA

*corresponding author: jxcheng@bu.edu; # equal contributions.

**Abstract:**
Mid-infrared photothermal microscopy is a new chemical imaging technology in which a visible beam senses the photothermal effect induced by a pulsed infrared laser. This technology provides infrared spectroscopic information at sub-micron spatial resolution and enables infrared spectroscopy and imaging of living cells and organisms. Yet, current mid-infrared photothermal imaging sensitivity suffers from a weak dependance of scattering on temperature and the image quality is vulnerable to the speckles caused by scattering. Here, we present a novel version of mid-infrared photothermal microscopy in which thermo-sensitive fluorescent probes are harnessed to sense the mid-infrared photothermal effect. The fluorescence intensity can be modulated at the level of 1% per Kelvin, which is 100 times larger than the modulation of scattering intensity. In addition, fluorescence emission is free of speckles, thus much improving the image quality. Moreover, fluorophores can target specific organelles or biomolecules, thus augmenting the specificity of photothermal imaging. Spectral fidelity is confirmed through fingerprinting a single bacterium. Finally, the photobleaching issue is successfully addressed through the development of a wide-field fluorescence-enhanced mid-infrared photothermal microscope which allows video rate bond-selective imaging of biological specimens.

**TOC Figure and Capture**

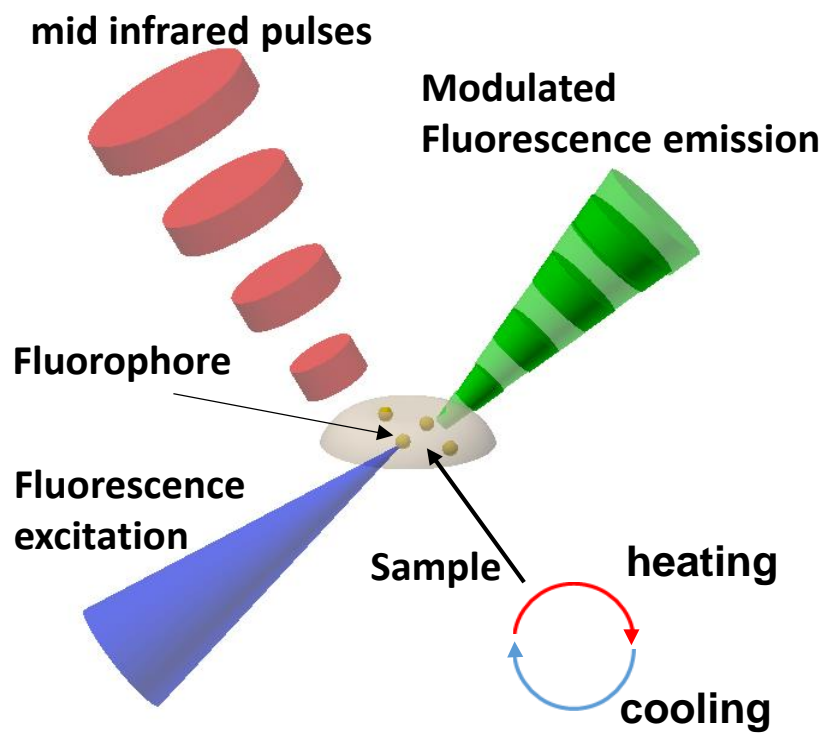

Modulation of fluorophore emission intensity by pulsed infrared pump of surrounding molecules.

**Introduction**
Visualizing the molecular composition and monitoring the molecular dynamics in a complex living system is a central theme of life science. Fluorescence microscopy has been widely adopted in biomedical research as it provides high speed background free imaging with exquisite molecular specificity[1-2] and superior resolution reaching the nanometer scale[3-5]. While fluorescence microscopy excels at mapping the distribution and dynamics of tagged organelles such as mitochondria and biomolecules such as glucose[6] and cholesterol[7], it does not provide chemical information of the tagged cells or organelles. Lacking such information hinders functional analysis, such as assessment of cell metabolic activity.

Providing chemical specificity, high speed and high sensitivity vibrational spectroscopic imaging is an emerging platform[8]. Recently developed coherent Raman scattering microscopy, based on coherent anti-Stokes Raman scattering (CARS) or stimulated Raman scattering (SRS), has allowed real-time vibrational imaging of biomolecules in living cells and tissues[9-10]. Advanced instrumentation has pushed the stimulated Raman spectral acquisition speed to microsecond scale[11]. Adoption of stable isotope probes and alkyne-based Raman tags greatly enhanced the detection sensitivity, specificity and functionality in SRS microscopy[10]. Being highly sensitive to C-H vibrations, CARS and SRS imaging have unveil new signatures of lipid metabolism in a variety of biological systems[12-14]. In comparison, high-speed CARS or SRS imaging of fingerprint Raman bands remains difficult.

Mid-infrared spectroscopy is complementary to Raman spectroscopy. Unlike Raman scattering, the infrared absorption cross section in the fingerprint region is larger than in the high-wavenumber C-H vibrational region. Fourier transform infrared (FTIR) spectroscopy is one of the most extensively used technique for chemical characterization and analysis of biological cells and tissues[15-18]. The inherent vibration absorption of mid-infrared photons by biological macromolecules including proteins, lipids, carbohydrates, and nucleic acids shows distinctive absorption bands. Shifts in relative heights of absorption bands, peak positions, and peak shape provide rich biomolecular information, including concentration, conformation, and orientation. FTIR spectroscopy has provided new insights in tissue classification[19], drug and tissue interaction[20], neurodegenerative diseases[21], cancer progression[22], and so on. However, the spatial resolution of infrared spectroscopic imaging is limited by the long mid-infrared illumination wavelength, ranging from 5 to 20 μm. Strong water absorption further hinders its application to living cells.

To overcome these limitations in infrared spectroscopy, a new platform, termed mid-infrared photothermal (MIP) microscopy, has been developed recently[23-30] to reach sub-micron spatial resolution[31]. The MIP effect relies on a photothermal process in which infrared absorption corresponds to a specific molecular vibrational bond causes a localized temperature rise at the vicinity of target molecules. This photothermal effect consequently induces a change of refractive index and a thermal expansion. The MIP signal is then obtained by probing these changes using a visible beam which provides a much smaller diffraction limit than the mid-infrared illumination, enabling a spatial resolution down to 300 nm[30] . Following the first demonstration of MIP imaging of living cells [23], technical innovations have been made to enable MIP detection in wide-field, using scattering or phase signals[32-35]. Meanwhile, MIP microscopy and its commercial product have found various applications in studying living cell, pharmaceuticals, viruses, and bacteria[25, 27-28, 36-40]. Yet, the photothermal effect induces only a tiny change in intensity and angular distribution of the scattered probe light, due to the weak thermal dependence of particle size and refractive index. Typical fractional change is on the order of $10^{-4}/K$, set by the intrinsic thermal properties of most materials. Such small modulation depth limits the signal to noise ratio (SNR), especially in the wide-field mode where the probe beam intensity at each pixel is limited by the well depth of a CMOS camera[33].

Here, we report a fluorescence enhanced MIP (FE-MIP) microscope that utilizes thermo-sensitive fluorescent dyes as probes of the photothermal effect. The FE-MIP principle is illustrated in **Figure 1a**. A sample stained with a thermo-sensitive dye is heated upon IR absorption by targeted molecules. The infrared pulse train heats the surrounding of the fluorescent probe and causes a temperature rise, which subsequently modulates the fluorescence emission efficiency. Such modulation is then measured by a lock-in amplifier. Using fluorescent dye to measure the temperature has been known[41-44]. Yet, it has not been used as a probe for infrared spectroscopic imaging. Our method offers a few key advantages over scattering-based MIP microscopy. The first is the utilization of much larger photothermal response of fluorescent dyes compared to scattering. Common fluorophores including FITC, Cy2, Cy3, Rhodamine, and green fluorescent protein have temperature-dependent emission efficiency on the order of $1\%/K$, which is nearly 100 times larger than scattering dependence on temperature. Thus, one can in principle boost the mid-infrared photothermal imaging speed by two orders of magnitude. Second, as fluorescence appears at a new wavelength from the incident beam, it is insensitive to laser relative intensity noise and the FE-MIP signal is intrinsically shot noise limited. Third, unlike scattering, fluorescence is not sensitive to speckles. Finally, fluorescent probes can target specific cells, intracellular organelles, or specific molecules, thus offering an enhanced specificity beyond the reach by scattering-based MIP microscopy. In this study, we report two FE-MIP systems, one in point-scanning mode and one in wide-field mode, as detailed below.

## Experimental Section

**Scanning FE-MIP microscope**: A pulsed mid-IR pump beam is generated by a tunable (from 1000 to 1886 cm$^{-1}$) quantum cascade laser (QCL, Daylight Solutions, MIRcat-2400) operating at 100 kHz repetition rate and 900 ns pulse duration. The IR beam passes through a calcium fluoride (CaF$_2$) cover glass and is then focused onto a sample through a gold-coating reflective objective lens (52×; numerical aperture (NA), 0.65; Edmund Optics, #66589). A continuous-wave probe laser (Cobolt, Samba) at 532 nm is focused onto the same spot from the opposite side by a refractive objective (60×; NA, 1.2; water immersion; Olympus, UPlanSApo). The probe beam is aligned to be collinear to the mid-IR pump beam. The reflective objective is fine tuned in 3D to ensure overlap of the two foci. A scanning piezo stage (Mad City Labs, Nano-Bio 2200) with a maximum scanning speed of 200 µs/pixel is used to scan the sample. The fluorescence is collected by the same refractive objective, reflected by a dichroic mirror (Thorlabs, DMSP550R, 550 nm cutoff), filtered by a long-pass filter (Thorlabs, FEL0550, 550 nm cut-off), and then collected by a PMT (Hamamatsu, H10721-110). Specimens on a CaF$_2$ coverglass are first imaged by fluorescence. Then, the pulsed IR laser is turned on and the modulated fluorescence signal is collected by the same PMT. The FE-MIP signal is extracted by a lock-in amplifier (Zurich Instruments, HF2LI). A laboratory-built resonant circuit is used to amplify the photocurrent from the PMT before it is sent to the lock-in. Before the reflective objective, the infrared laser passes through the CaF$_2$ slip and the reflected infrared laser intensity is measured by a mercury cadmium telluride (MCT) detector for normalization of IR power at each wavelength.

**Wide-field FE-MIP microscope:** The IR pulses are generated by the same QCL used in scanning FE-MIP. The visible probe beam for fluorescence excitation (wavelength at 488 nm or 520 nm) is obtained by second-harmonic generation of a quasi-continuous femtosecond laser tuned to 976 nm or 1040 nm (Coherent Inc, Chameleon, 140 fs, 80 MHz). Prior to second-harmonic generation, the femtosecond beam is chopped into a 200-kHz pulse train (300 ns pulse width) by an acousto-optical modulator (AOM, Gooch and Housego). The IR beam passes through the substrate and is weakly focused onto a sample by a parabolic mirror (f = 15 mm, Thorlabs, MPD00M9-M01). Using a Kohler illumination configuration, the probe beam is focused on the back focal plane of the objective lens (50×, 0.8 NA, Nikon) by a condenser (f = 75 mm, AC254-075-A, Thorlabs). The fluorescence emission is collected by the same objective lens, and after a long pass filter, collected by a CMOS camera (FLIR, Grasshopper3 GS3-U3-51S5M). The FE-MIP images are acquired by a virtual lock-in camera approach [33]. Briefly, a pulse generator (Emerald Pulse Generator, 9254-TZ50-US, Quantum Composers) generates a master clock signal at 200 kHz and externally triggers the QCL, the AOM and the CMOS camera to synchronize the IR pump pulses, the probe pulses, and camera exposure.

**Cancer cell culture and staining:** Mia Paca2 cells were purchased from the American Type Culture Collection (ATCC). The cells were cultured in RPMI 1640 medium supplemented with

10% FBS and 1% P/S. All cells were maintained at 37°C in a humidified incubator with 5% $CO_2$ supply. For Nile red staining, cells were incubated with 10µM Nile Red (Invitrogen) for 30 minutes at room temperature followed with 15 minutes fixation in 10% neutral buffered formalin. For rhodamine 123(Invitrogen) staining, cells were incubated with 10 µg/ml rhodamine 123 for 30 minutes at 37°C.

**Bacterial culture and staining:** Staphylococcus aureus (S. aureus) was incubated in a MHB medium for 10 h. After centrifuging and washing in phosphate-buffered saline (PBS). The bacteria were fixed by formalin solution for 0.5 h. Rhodamine 6G or Cy2 at $10^{-4}$ M was then added into the bacteria pellet. The pellet was then resuspended and incubated for 1 h. With final washing steps, 2 µL sample were dried on a $CaF_2$ coverslip for imaging. Shigella flexneri expressing GFP was grown overnight at 37°C on a tryptic soy agar plate. Colonies with fluorescent green were picked up by sterile inoculation loops and then resuspended in PBS. The bacterial solution was diluted by optical density at 600 nm (OD600) to 0.1. The bacteria were then fixed by 10% Formalin for 30 min at room temperature. The bacteria solution was washed twice by PBS before imaging.

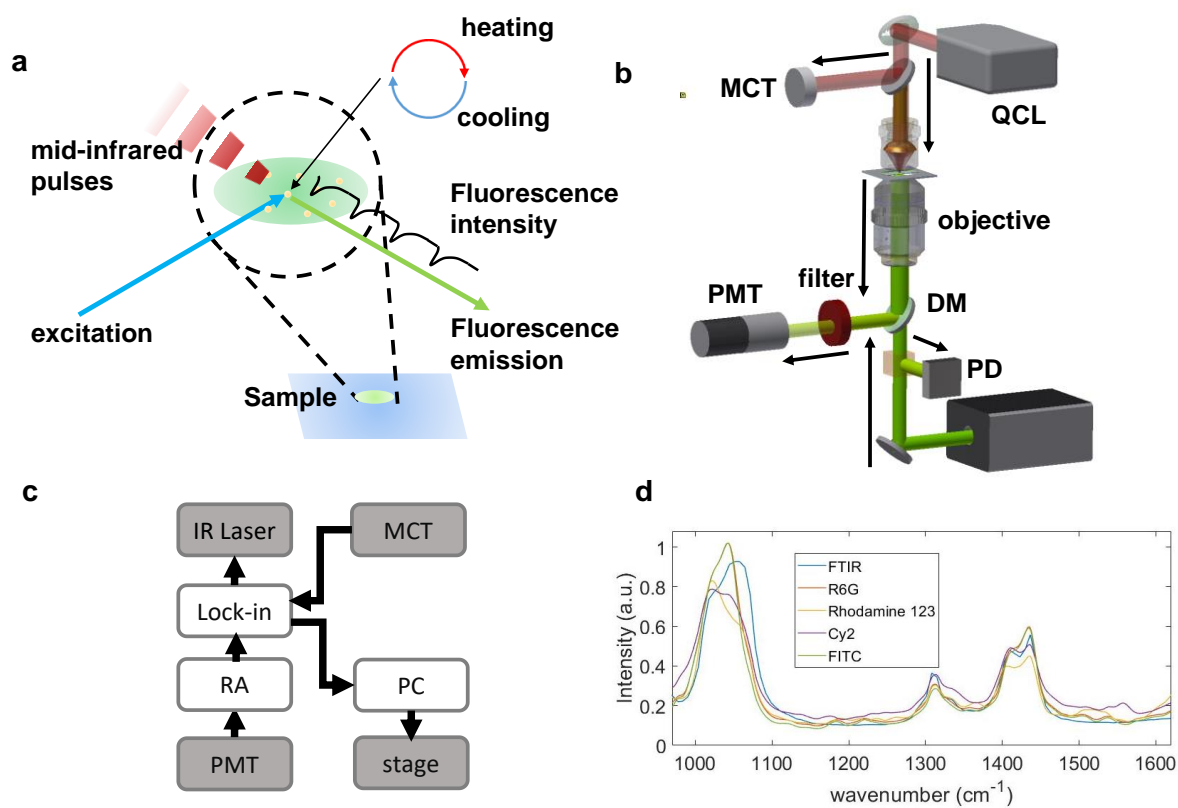

**Figure 1. Fluorescence enhanced mid-infrared photothermal (FE-MIP) sensing principle, point scanning microscope, and spectral fidelity**. (a) Modulation of fluorophore emission intensity by pulsed infrared pump of surrounding molecules. (b) Schematic of a point scanning FE-MIP microscope. A pulsed mid-IR pump beam from a quantum cascade laser (QCL) and a continuous visible fluorescence excitation beam are focused at the sample with a reflective objective and a water-immersion objective, respectively. The fluorescence emission is reflected by a dichroic mirror (DM), filtered and directed to photomultiplier tube (PMT). A beam splitter is placed to reflect the scattered visible beam to a photodiode for scattering-based MIP (Sc-MIP) imaging. (c) Electronics connection. The photothermal signal is detected by a PMT connected to a resonant amplifier (RA) and detected by a lock-in amplifier. A PC is used for controlling the scanning stage and data acquisition. (d) FE-MIP

spectra of DMSO supplemented with various thermo-sensitive fluorescent dyes. The FTIR spectrum of DMSO is shown for comparison.

## Results and Discussion
### Point-scanning FE-MIP microscope and spectral fidelity

Based on the principle shown in **Figure 1a**, we have built a scanning FE-MIP microscope as shown in **Figure 1b**. A QCL laser provides IR pulses tunable in the entire fingerprint region. The repetition rate and pulse width were set to be 100 kHz and 900 nanoseconds, respectively. Two CW lasers at 532 nm and 488 nm were used for fluorescence excitation. The fluorescence was detected by a photomultiplier tube and the FE-MIP signal was extracted by a lock-in amplifier. On the same setup, a photodiode was also installed for scattering-based MIP (Sc-MIP) imaging using the 532-nm laser as the probe beam. The electronics connections are shown in **Figure 1c**.

Using this system, we have validated the spectral fidelity of FE-MIP microscope. We dissolved various thermo-sensitive dyes in DMSO, at 100 µM concentration. We then recorded the FE-MIP signals while scanning the QCL laser. At each wavenumber, the FE-MIP intensity was normalized by the IR intensity measured by MCT. In all cases (**Figure 1d**), the FE-MIP spectra show the same peak intensity and width as the FITR spectrum of DMSO. Importantly, because the dye concentration (100 µM) is much lower than DMSO concentration (14 M), the dyes do not interfere with the FE-MIP spectra. These data demonstrate the FE-MIP microscope is able to produce reliable spectral information.

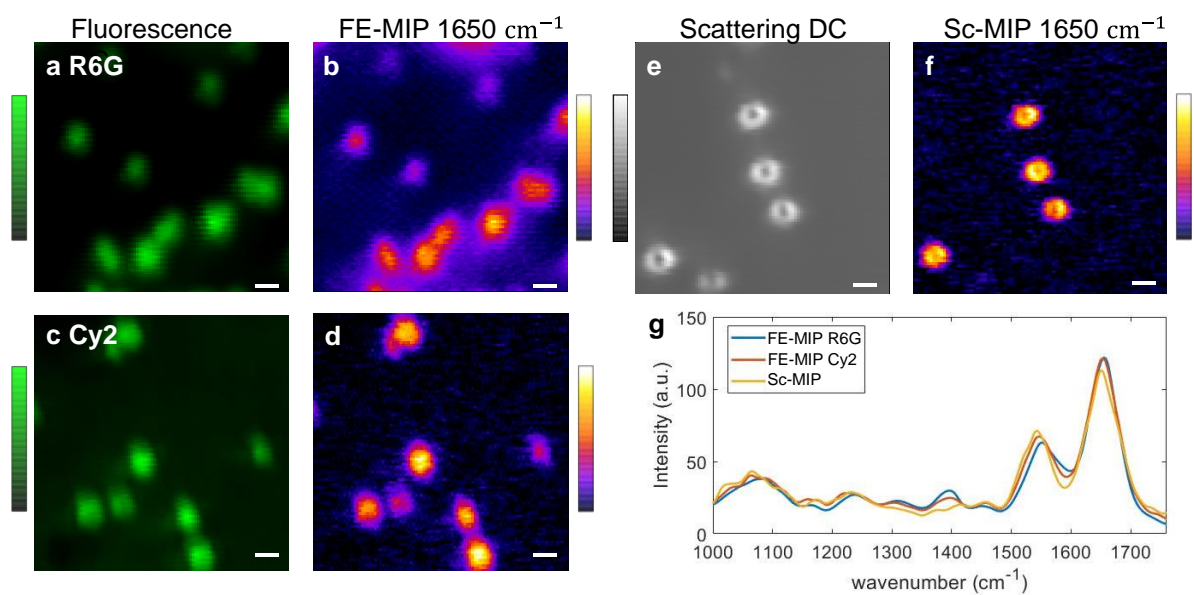

**Figure 2. FE-MIP imaging and fingerprinting single S. aureus.** (a) Fluorescence image of S. aureus stained with R6G. Scale bar: 1 $\mu m$. Pixel dwell time: 1 ms. Fluorescence excitation: 0.025 mW at 532 nm. (b) FE-MIP image of same S. aureus at 1650 cm$^{-1}$. (c) Fluorescence image of S. aureus stained with Cy2. Fluorescence excitation: 0.025 mW at 488 nm. (d) FE-MIP image of same S. aureus at 1650 cm$^{-1}$. (e) Scattering image of S. aureus. Pixel dwell time: 1 ms. Visible probe power: 1.5 mW. (f) Sc-MIP image of same S. aureus at 1650 cm$^{-1}$. (g) The fingerprint spectra of S. aureus measured by FE-MIP and Sc-MIP, respectively. For both FE-MIP and Sc-MIP, the IR laser power 1650 cm$^{-1}$ was 10.2 mW at sample.

### FE-MIP imaging and fingerprinting of single bacteria

We applied the FE-MIP microscope to image single S. aureus bacteria to evaluate its chemical imaging capability on biological specimens (**Figure 2**). The S. aureus culture was diluted to a concentration of around 5×10$^5$/mL and then dried on a CaF$_2$ substrate. The S. aureus particles were stained with fluorescence dye Cy2 and R6G, respectively. For each specimen, we acquired fluorescence and FE-MIP images of the same bacteria at 1650 cm$^{-1}$ targeting the amide I band. For the FE-MIP images, SNRs of 26 and 34 were achieved for R6G and Cy2 labeled bacteria at the fluorescence excitation power of 0.025 mW at the sample. For comparison, we recorded scattering and MIP images of the same specimen. In order to obtain a similar SNR of 37, an excitation power of 1.5 mW at sample was required, which is 60 times of the probe power used for FE-MIP. On a single bacterium, we recorded the vibrational fingerprint spectrum (**Figure 2g**). The FE-MIP spectra based on R6G and Cy2 matches well the Sc-MIP spectrum, showing distinct peaks at 1650, 1550, and 1080 cm$^{-1}$ for protein amide I, protein amide II, and nuclei acid phosphate vibrations, respectively. Notably, the Sc-MIP image shows a ring structure, with bright contrast from the peripheral of the cell, whereas the FE-MIP imaged shows bright contrast from the entire cell. This result indicates that FE-MIP is immune to interferences encountered in the scattering image.

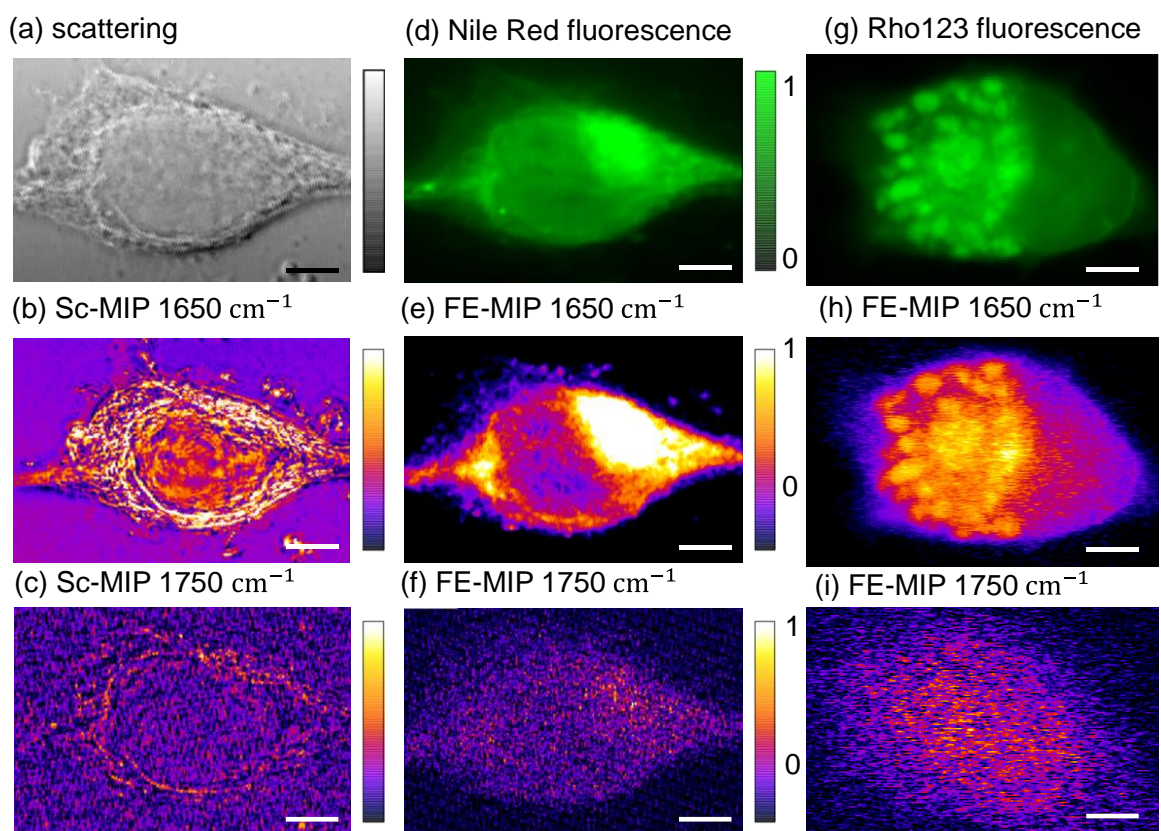

**Figure 3. Sc-MIP and FE-MIP images of living MiaPaca2 cancer cells.** (a) Scattering image of a MiaPaca2 cell. (b) Sc-MIP image of the same cell at 1650 cm$^{-1}$. (c) Sc-MIP image at 1750 cm$^{-1}$. (d) Fluorescence image of the same MiaPaca2 cell stained with Nile red. The probe laser is at 532 nm. (e) FE-MIP image at 1650 cm$^{-1}$. (f) FE-MIP image at 1750 cm$^{-1}$. (g) Fluorescence image of a different MiaPaca2 cell stained with Rhodamine 123 (Rho123). The probe laser is at 488 nm. (h) FE-MIP image at 1650 cm$^{-1}$. (i) FE-MIP image at 1750 cm$^{-1}$. Scale bar: 5 μm.

**SC-MIP and FE-MIP imaging of specific organelles inside a cancer cell**
Compared to bacteria, eukaryotic cells contain a nucleus and highly organized organelles in the cytoplasm. In the transmission image shown in **Figure 3a**, the scattering-based contrast shows the overall cell morphology. Accordingly, the SC-MIP image (**Figure 3b**) at 1650 cm$^{-1}$ corresponding to the amide I band shows the protein content inside the nucleus and protein-rich structures in the cytoplasm, without specificity to a certain organelle. In contrast, fluorescence microscopy is able to visualize specific biomolecules and/or intracellular organelles via the versatile fluorescent probes. For instance, Nile red can selectively stain the intracellular lipid droplets and membranes[45] while Rhodamine 123 is a specific probe for localizing mitochondria in living cells[46]. By staining the same cell with Nile red and excitation of the dye at 532 nm, phospholipid membranes in the cell are visualized (**Figure 3d**), where the brightest contrast is likely from the ER membrane. The FE-MIP image at 1650 cm$^{-1}$ gives the distribution of proteins associated with the membranes labeled by Nile red (**Figure 3e**). The contrast nearly disappears when the IR laser is tuned to 1750 cm$^{-1}$ (**Figure 3f**), showing the chemical specificity. To show that our method is applicable to other organelles, we performed an independent experiment in which living MiaPaca2 cells were labeled by Rhodamine 123 targeting intracellular mitochondria (**Figure 3g**). In accordance, the FE-MIP image at 1650 cm$^{-1}$ shows selective and bright contrast from the mitochondria (**Figure 3h**) and the contrast disappears at 1750 cm$^{-1}$ (**Figure 3i**).

## From point-scanning to wide-field FE-MIP

In the above experiments, the power used for FE-MIP imaging is at the microwatt level and is 60 times less than the power used for scattering MIP imaging. The extremely low photon budget for FE-MIP imaging opens the opportunity of increasing the throughput via wide-field illumination with IR pump pulses and visible probe pulses. Importantly, compared to scanning FE-MIP, wide-field FE-MIP could significantly reduce the fluorophore photobleaching rate based on the following consideration. In our experiment, the IR pulse is 900 ns in duration and the pulse-to-pulse duration is 10 μs. In the aqueous environment, the temperature profile largely follows the IR pulse. Thus, in a scanning experiment where a continuous wave probe laser is used, the duty cycle is about 10%. Yet, in a wide field measurement, we only need two visible pulses to measure the IR-on and IR-off states. Thus, the duty cycle can be 50%. In this way, the probe laser power can be much reduced, thus alleviating the photobleaching issue. Notably, our group recently demonstrated scattering-based wide-field MIP imaging[33]. Yet, the signal to noise ratio at the high-speed mode is limited by the weak dependence of scattering on temperature and the small well depth of the CMOS camera. As a result, a large number of integrations were needed to accumulate sufficient photons to probe the MIP signal. Unlike the scattering photons, the fluorescence usually does not saturate the camera. Due to the high thermo-sensitivity of fluorescent probes, it is anticipated that the MIP signal can be extracted from two sequential frames (hot and cold) without further average.

## Wide field FE-MIP imaging system

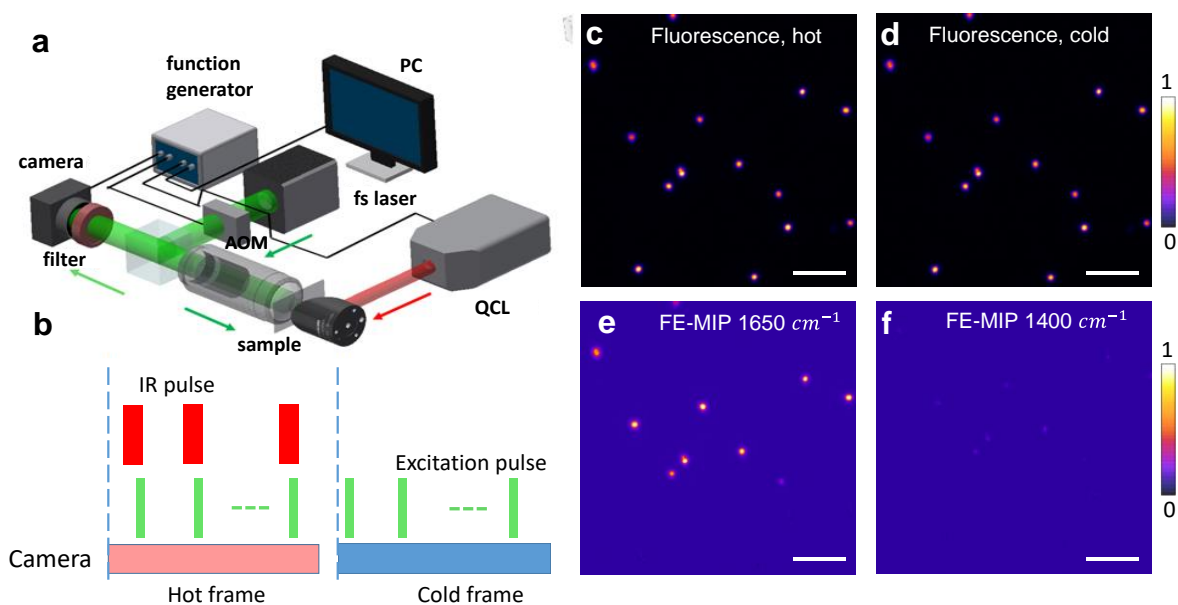

**Figure 4. The wide field FE-MIP microscope.** (a) Schematic of the wide field FE-MIP microscope. The infrared laser generated by the QCL is focused by a parabolic mirror. The visible light modulated by an AOM illuminates the sample and excites the fluorescence. The fluorescence is filtered and collected by a CMOS camera. (b) Temporal synchronization of IR pump pulse, fluorescence excitation

pulse, and camera exposure. (c,d) Wide field fluorescence image of Cy2-stained S. aureus with IR on and IR off, designated as hot and cold, respectively. (e) Wide field FE-MIP at 1650 cm$^{-1}$ (f) FE-MIP at 1400 cm$^{-1}$ off resonance. Scale bar: 10 μm. Fluorescence excitation: 488 nm, 1 mW. IR pulse rate: 200 KHz. Camera frame rate: 40 Hz.

Based on the above rationale, we built a wide field FE-MIP microscope as shown in **Figure 4a**. The pulsed Infrared laser is weakly focused onto a sample by a parabolic mirror. An 80-MHz femtosecond laser is modulated by an AOM and frequency-doubled to visible window. The broad bandwidth of femtosecond pulses reduces speckles in scattering-based MIP imaging. The virtual lock-in detection scheme is illustrated in **Figure 4b**. The fluorescence excitation pulse is synchronized with the IR pump pulse. In the cold frame, no IR pump pulses heat the sample. The camera detects the hot and cold frames sequentially at 40 Hz frame rate. The difference (cold − hot) generates the MIP image. To characterize the spatial resolution, we mapped Cy2-label polystyrene beads with diameter of 500 nm, the FE-MIP intensity profile shows a full-width at half maxim of 610 nm. After deconvolution with particle size, the spatial resolution is estimated to be 350 nm, which is close to the diffraction limit of the 0.8 NA objective.

To demonstrate the applicability of wide-field FE-MIP to biological specimens, we deposited S. aureus stained with Cy2 onto a silicon substrate and measured the fluorescence with IR on and IR off sequentially. The hot frame and the cold frame are illustrated in **Figure 4c** and **d**, respectively. By subtracting the hot from the cold frame, the intensity difference generates the FE-MIP image shown in **Figure 4e**. When the IR laser is tuned to 1650 cm$^{-1}$, corresponding to the amide I band of proteins, a signal to noise ratio of 30 was obtained. The IR pulse only heats the upper half of the field of view. For this reason, only the S. aureus particles in the upper part give the FE-MIP contrast. When the IR laser is tuned to 1400 cm$^{-1}$, off resonance to major IR peaks, the contrast nearly disappears (**Figure 4f**). We note that the wide-field FE-MIP imaging speed of 20 frames per second is 2000 times faster than the scanning FE-MIP imaging speed (100 seconds per frame with a pixel dwell time of 1.0 millisecond).

**Performance comparison between wide-field Sc-MIP and FE-MIP**
Next, we compared the performance of fluorescence-enhanced and scattering-based MIP imaging in the wide field mode, using S. aureus on a silicon substrate as testbed. **Figure 5a-c** shows the fluorescence and FE-MIP images of individual S. aureus with the IR laser tuned to 1650 cm$^{-1}$. The signal to noise ratio for a single bacterium reaches 33 in single frame FE-MIP and 275 after 100 frames average. **Figure 5d-f** shows the scattering and SC-MIP images of individual S. aureus with the IR laser tuned to 1650 cm$^{-1}$. **Figure 5g-h** shows the intensity profile across the white line indicated in the images. The Sc-MIP contrast is completely buried in the speckle pattern in singe frame acquisition. After 100 frames average, the signal to noise ratio reaches 14, which is 20 times lower than FE-MIP of the same sample. The much higher signal to noise ratio in FE-MIP can be attributed to the lack of shot noise from scattered photons and the much larger thermo-sensitivity of the fluorescence probe. **Figure 5h** shows similar line width for the bacterium in Sc-MIP and FE-MIP. However, the FE-MIP intensity does not suffer from the interference (i.e. the dark ring around the peak) encountered in the Sc-MIP image. **Figure 5i-j** shows the FE-MIP and Sc-MIP intensities from a single bacterium in sequentially acquired hot and cold frames. The modulation depth, defined as percentage of intensity difference between hot and cold frames, is found to be about 4% for FE-MIP, but is buried in frame to frame fluctuations in Sc-MIP. Moreover, **Figure 5i** shows negligible photobleaching in the recorded 50 frames over a period of 1.25 seconds.

Besides fluorescent dyes, we tested the feasibility of FE-MIP imaging for cells expressing green fluorescent proteins (GFPs). It has been shown that GFPs are highly thermo-sensitive with 1% intensity decrease per degree in temperature rise [47-48]. Accordingly, in FE-MIP imaging of Sheila Flexneri bacteria expressing GFP (**Figure 5k-m**), we observed 2% fluorescence intensity difference between cold and hot frames (**Figure 5n**). The signal to noise ratio reaches 10 and 97 in single frame and 100 frames average, respectively. The recorded 50 frames only experienced 3% photobleaching. The fluorescence fluctuation from frame to frame was due to laser instability.

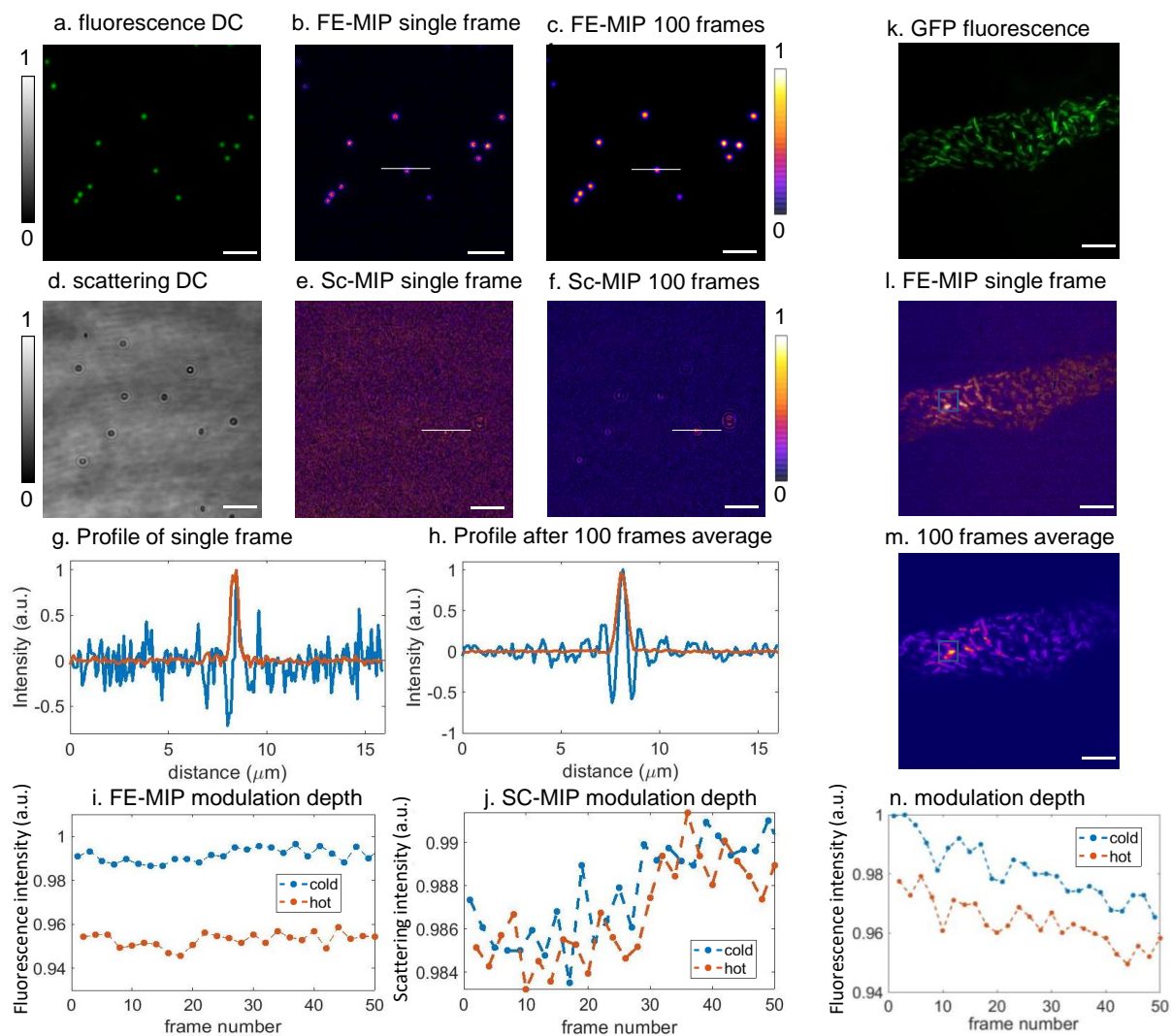

**Figure 5. Performance comparison between wide field SC-MIP and wide field FE-MIP systems.** (a) Fluorescence image of S. aureus deposited on a silicon substrate. (b) Single frame FE-MIP image of the same cells at 20 Hz speed. (c) FE-MIP image of the same cells with 100 frames average. (d) Scattering image of S. aureus deposited on a silicon substrate. (e) Single frame Sc-MIP of the same cells shown in panel d at 20 Hz speed. (f) Sc-MIP of the same field of view with 100 frames average. (g) Intensity profile along the white line marked in panel b and e. (h) Intensity profile along the white line marked in panel c and f. (i) Fluorescence intensity of single S. aureus in sequentially acquired hot and cold frames. (j) Scattering intensity of single S. aureus in sequentially acquired hot and cold frames. The IR laser is tuned to 1650 cm$^{-1}$ for data in panel a to h. (k) Fluorescence image of Sheila Flexneri bacteria expressing GFP. (l) Single frame FE-MIP of the same bacteria at 1650 cm$^{-1}$. (m) FE-MIP image of the same field of view with 100 frames average. (n) Fluorescence intensity of the square area in sequentially acquired hot and cold frames.

Recently, the Tokmakoff and Min groups reported two related methods that code vibrational signatures into fluorescence spectroscopy, based on femtosecond-scale infrared pulse excitation [49] and picosecond-scale stimulated Raman excitation [50], respectively. These methods are based on the up-conversion process, in which vibrational excitation and fluorescence emission are from the same dye molecule. Our FE-MIP method is fundamentally different in that fluorescence emission efficiency from a thermo-sensitive probe is modulated by nanosecond-scale pulsed infrared excitation of molecules of interest surrounding the probe.

Theoretically, FE-MIP microscopy is based the thermal diffusion from the target molecule to the fluorescent probe. The thermal diffusion length is defined as $\mu_t = 2\sqrt{\alpha t}$, where α is the thermal diffusivity[51-52]. In aqueous environment, the value of α is $1.4 \times 10^{-7}$ m$^2$/s [53]. In our wide-field FE-MIP experiment, the IR pulse is 900 ns in duration and the fluorescence excitation pulse is 300 nm. If we set the pump-probe delay to be 900 ns, the thermal diffusion length is ~700 nm, which is slightly larger than the diffraction limit of the visible probe beam. If one can use IR pump and visible probe pulse of 5 ns duration and the pump-probe delay is set to be 5 ns, the thermal diffusion length can be reduced to 50 nanometer. In this case, one can detect the chemical content surrounding the fluorescence probe on the nanoscale.

## Conclusions
In efforts to push the detection limit and increase the specificity of optically detected mid-infrared photothermal microscopy, a new platform termed fluorescence-enhanced mid-infrared photothermal (FE-MIP) microscopy is developed. Our platform harnesses thermo-sensitive fluorescent probes to sense surrounding temperature rise induced by pulsed infrared excitation. High spectral fidelity is demonstrated for fluorescent probes in DMSO solution and inside biological cells. In the point scanning modality, we have demonstrated FE-MIP imaging and fingerprinting of a single bacterium. While using fluorescence as a read out, the fingerprint information would allow functional assessment of biological specimen, such as metabolic response of bacteria to antibiotics treatment. Furthermore, organelle-specific FE-MIP imaging is achieved, which opens exciting opportunities of probing the chemical content of intracellular organelles. In the wide-field modality, we demonstrated video rate, high signal to noise ratio, speckle-free FE-MIP imaging of individual bacteria. Finally, our platform is applicable to biological cells expressing GFP. This approach opens new opportunities of monitoring secondary structure of specific proteins tagged by GFP, which is beyond the reach by IR spectroscopy or fluorescence spectroscopy alone.

**Acknowledgements**: This work is supported by R35GM136223 to JXC.